\documentclass[superscriptaddress,showpacs,preprintnumbers,amsmath,amssymb,twocolumn,floats]{revtex4}
\usepackage{txfonts}
\usepackage{amssymb}
\usepackage{graphicx}
\usepackage{sidecap}
\usepackage{CJK}
\usepackage{txfonts}

\begin{document}

\title{Bulk and surface electronic structure of trigonal structured PtBi$_2$ studied by angle-resolved photoemission spectroscopy}

\author{Q. Yao}
\affiliation{State Key Laboratory of Surface Physics, Department of Physics, and Laboratory of Advanced Materials, Fudan University, Shanghai 200433, China}
\affiliation{State Key Laboratory of Functional Materials for Informatics, Shanghai Institute of Microsystem and Information Technology (SIMIT), Chinese Academy of Sciences, Shanghai 200050, China}

\author{Y. P. Du}
\affiliation{National Laboratory of Solid State Microstructures, College of Physics, Nanjing University, Nanjing 210093, China}
\affiliation{Collaborative Innovation Centre of Advanced Microstructures, Nanjing 210093, China}

\author{X. J. Yang}
\author{Y. Zheng}
\affiliation{Department of Physics, Zhejiang University, Hangzhou 310027, China}

\author{D. F. Xu}

\author{X. H. Niu}

\author{X. P. Shen}
\affiliation{State Key Laboratory of Surface Physics, Department of Physics, and Laboratory of Advanced Materials, Fudan University, Shanghai 200433, China}

\author{H. F. Yang}
\affiliation{State Key Laboratory of Functional Materials for Informatics, Shanghai Institute of Microsystem and Information Technology (SIMIT), Chinese Academy of Sciences, Shanghai 200050, China}

\author{P. Dudin}

\author{T. K. Kim}

\author{M. Hoesch}
\affiliation{Diamond Light Source, Harwell Science and Innovation Campus, Didcot OX11 0DE, United Kingdom}

\author{I. Vobornik}
\affiliation{CNR-IOM, TASC Laboratory AREA Science Park-Basovizza, 34149 Trieste, Italy}

\author{Z. -A. Xu}
\affiliation{Collaborative Innovation Centre of Advanced Microstructures, Nanjing 210093, China}
\affiliation{Department of Physics, Zhejiang University, Hangzhou 310027, China}
\affiliation{Zhejiang California International NanoSystems Institute, Zhejiang University, Hangzhou 310027, China}

\author{X. G. Wan}
\affiliation{National Laboratory of Solid State Microstructures, College of Physics, Nanjing University, Nanjing 210093, China}
\affiliation{Collaborative Innovation Centre of Advanced Microstructures, Nanjing 210093, China}

\author{D. L. Feng}
\affiliation{State Key Laboratory of Surface Physics, Department of Physics, and Laboratory of Advanced Materials, Fudan University, Shanghai 200433, China}
\affiliation{Collaborative Innovation Centre of Advanced Microstructures, Nanjing 210093, China}

%\end{CJK*}
\author{D. W. Shen}
\email{dwshen@mail.sim.ac.cn}
\affiliation{State Key Laboratory of Functional Materials for Informatics, Shanghai Institute of Microsystem and Information Technology (SIMIT), Chinese Academy of Sciences, Shanghai 200050, China}
\affiliation{CAS Center for Excellence in Superconducting Electronics (CENSE), Shanghai 200050, China}

\begin{abstract}

PtBi$_2$ with a layered trigonal crystal structure was recently reported to exhibit an unconventional large linear magnetoresistance, while the mechanism involved is still elusive. Using high resolution angle-resolved photoemission spectroscopy, we present a systematic study on its bulk and surface electronic structure. Through careful comparison with first-principle calculations, our experiment distinguishes the low-lying bulk bands from entangled surface states, allowing the estimation of the real stoichiometry of samples. We find significant electron doping in PtBi$_2$, implying a substantial Bi deficiency induced disorder therein. We discover a Dirac-cone-like surface state on the boundary of the Brillouin zone, which is identified as an accidental Dirac band without topological protection. Our findings exclude quantum-limit-induced linear band dispersion as the cause of the unconventional large linear magnetoresistance.

\end{abstract}


\maketitle

%---------------------------introduction-------------------------

\section{Introduction}

Ordinary magnetoresistance (MR), usually a weak effect of only a few percent change of the electrical resistance in response to an applied magnetic field, is a common property of nonmagnetic materials. It is quadratic in the external magnetic field \textbf{B} at low fields, then saturates at high fields. Recently, with the discovery of the anomalous large MR effects in graphene, topological insulators, Dirac semimetals and some other materials~\cite{graphene, Bi2Se3, Cd3As2 MR, WTe2 MR, Bi thin films MR, Ag2Se Ag2Te MR}, there has been increased interest in searching for more materials with a large MR effect due to its potential applications~\cite{introduction1, introduction2}. Several scenarios have been proposed to explain the large MR. For example, a linear MR has been suggested to be closely linked with the quantum limit induced by Dirac-like band dispersion, disorders in materials while disorders with high mobility or hole-electron compensation usually contribute to the large MR~\cite{MR quantum model, MR quantum model1, MR quantum model2, MR classical model, Jiang WTe2 PRL}.

Recently, PtBi$_2$ with a layered trigonal crystal structure was reported to exhibit a large linear MR up to 682\% under B~=~15~T at T~=~2~K~\cite{Zhuan}. Further transport measurements reveal that the slope of the MR scales well with the Hall mobility, implying that its giant linear MR might be dominated by disorder but is unlikely to be accounted for the linear energy dispersion induced by the quantum limit as in topological insulators, graphene and Dirac semimetals~\cite{graphene, Bi2Se3, Cd3As2 MR, WTe2 MR}. Nevertheless, considering the complexity of MR effect, transport measurements can not completely rule out other band structure factors that may contribute to the anomalous MR behavior of PtBi$_2$. Thus a comprehensive understanding of the electronic structure of PtBi$_2$ is a prerequisite to ascertaining the origin of the anomalous transport properties~\cite{WTe2 MR, Jiang WTe2 PRL}.

In this article, we report a comprehensive picture of the low-lying surface and bulk electronic structure of the trigonal PtBi$_2$, which is achieved by combining angle-resolved photoemission spectroscopy (ARPES) measurements with surface and bulk band structure calculations. We identify the entangled bulk and surface states, enabling us to estimate the real stoichiometry of PtBi$_2$ and resolve a Dirac-cone-like surface state on the boundary of the Brillouin zone (BZ) through careful comparison with calculations. We discover significant electron self-doping in this material, indicative of a substantial Bi deficiency, which is consistent with our electron probe X-ray microanalysis (EPMA) measurements. We associate the accidental surface Dirac cone band with the symmetry requirement at the $M$($L$) point of the BZ, which is not protected topologically and would not contribute much to the anomalously large linear MR. Our findings provide evidence that the classical disorder model accounts for the unconventional large linear MR~\cite{MR classical model, MR classical model2}.

%-----------------------experiment and results---------------------

\section{Experimental}

\begin{figure}[t]
\centering
\includegraphics[width=8.5cm]{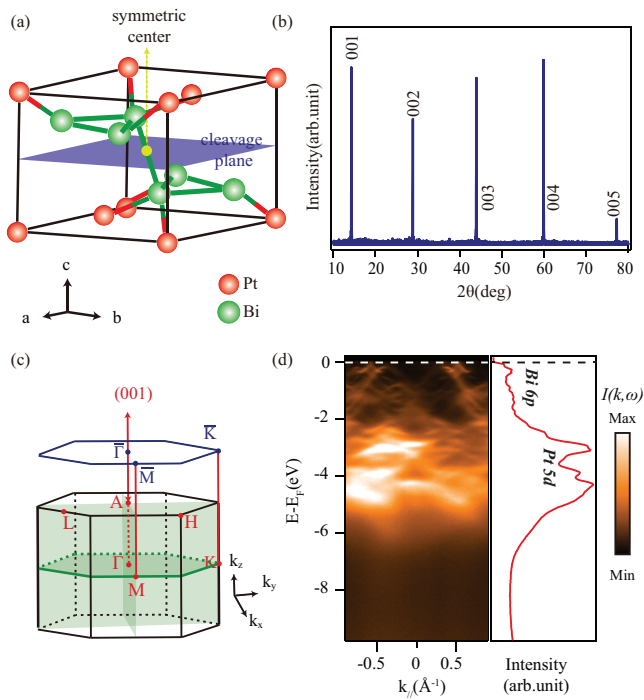}
\caption{(Color online) (a) Sketch of the trigonal structure. Red spheres represent Pt atoms and green spheres represent Bi atoms. The symmetric center is represented by the yellow point. The blue dashed line represents the cleaved plane between the two Bi layers. (b) X-ray diffraction pattern of PtBi$_2$ single crystal with a series of (00n) peaks. (c) The three-dimension Brillouin zone of PtBi$_2$. (d) The valence band structure around the $A$ point along the $L$-$A$-$L$ direction and corresponding angle-integrated energy distribution curves (EDCs).}
\label{lattice}
\end{figure}

High-quality single crystals of layered trigonal PtBi$_2$ were synthesized by a self-flux method as described elsewhere~\cite{Zhuan}. Powder X-ray diffraction (XRD) was performed with a BRUKER/AXS D8 Advance X-ray Diffractometer using Cu K$\alpha$ radiation from 10$^\circ$ to 80$^\circ$ with a step of 0.02$^\circ$, and the sample composition was measured by an electron-probe microanalysis (Shimadzu, EPMA-1720) at room temperature. The ARPES measurements were performed with photons from both the APE beamline of Elettra Synchrotron Trieste and the I05 beamline of Diamond light source. The beamlines are equipped with VG-Scienta DA30 and R4000 electron analyzers, respectively. The overall energy resolution was 10$\sim$15~meV depending on the photon energy, and the angular resolution was 0.3$^\circ$. The samples were cleaved \emph{in situ} under a vacuum better than 5 $\times$10$^{-11}$ Torr and measured at 10~K.

\begin{figure*}[t]
    \centering
    \includegraphics[width=17.5cm]{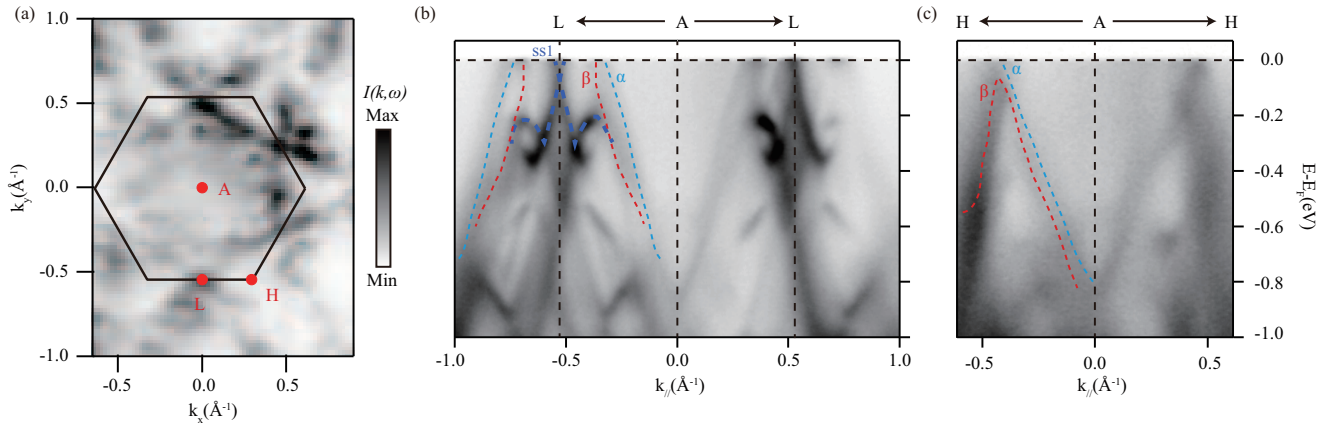}
    \caption{(Color online) (a) Photoemission intensity map integrated over [$E_F$ $-$ 5meV, $E_F$ $+$ 5meV]. The hexagon is the BZ. (b), (c) The photoemission intensity plots along the \emph{L}-\emph{A}-\emph{L} and \emph{H}-\emph{A}-\emph{H} directions, respectively. Data were taken at 10K with circularly polarized photons of h$\nu$~=~64~eV.}
\label{dispersion}
\end{figure*}

Our first principles calculation is performed using the VASP (Vienna ab initio simulation package) code~\cite{theory1,theory2}. The results are obtained by using the generalized gradient approximation (GGA) Perdew-Becke-Erzenhof (PBE)~\cite{theory3} function. An energy cutoff of 400 eV was adopted for the plane-wave expansion of the electronic wave function and the energy convergence criteria was set to 10$^{-5}$ eV. Appropriate k-point meshes of (11$\times$11$\times$1) and (100$\times$100$\times$1) were used for self-consistent and Fermi surface calculations, respectively. The spin-orbit coupling was taken into account by the second variation method~\cite{theory4}. To simulate the (001) surface of PtBi$_2$, we built a slab model composed of nine unit cells, in which both of the top and bottom surfaces are terminated by Bi atoms. A vacuum spacing of 20 $\AA$ was used so that the interaction in the non-periodic directions could be neglected. The experimental lattice parameters were used~\cite{PtBi2 crystal}. We projected the band structure and Fermi surface to the first unit cell to get the surface electronic structure which fits the experimental band structure very well.

\section{Results and Discussions}

%Fig3 theoretical calculations with experimental results
\begin{figure}[t]
    \includegraphics[width=8.5cm]{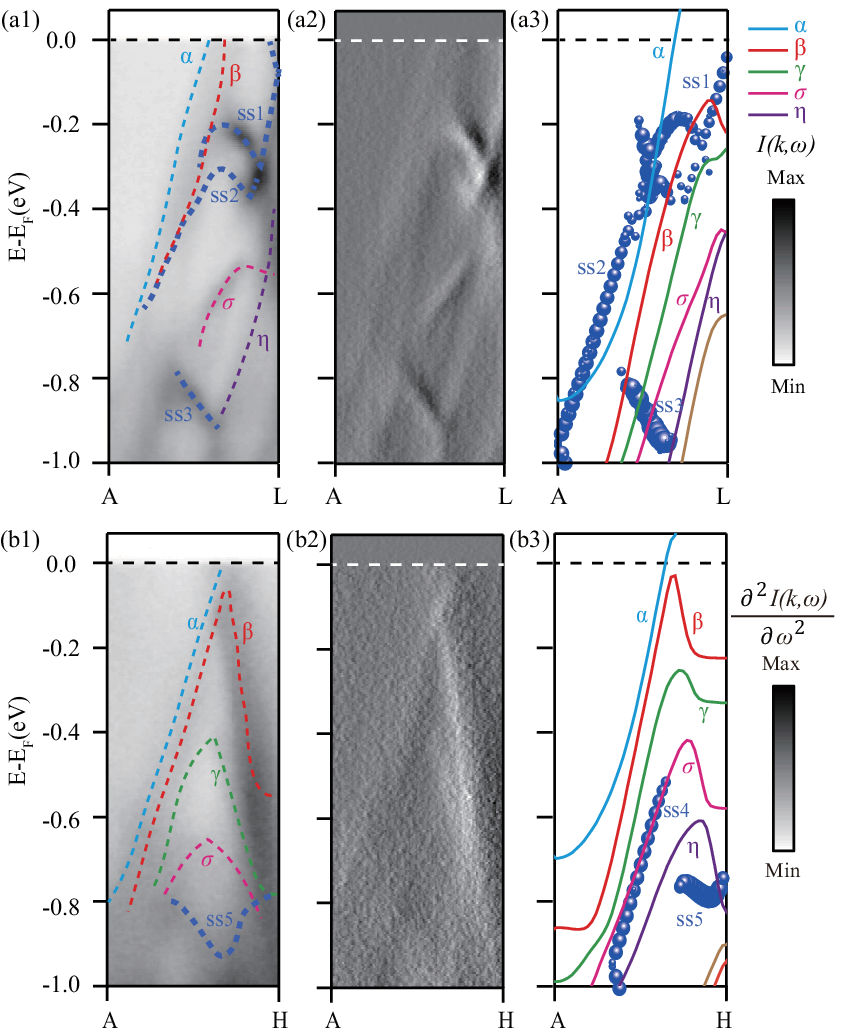}
    \caption{(Color online)  (a1-a3) and (b1-b3) The photoemission intensity plots, the second derivative and corresponding theoretical calculations along \emph{A}-\emph{L}-\emph{A} and \emph{H}-\emph{A}-\emph{H} directions, respectively. Blue spheres are calculated surface states for a nine-unit-cell-thick (001) slab with Bi terminations on the top surface. The size of the blue spheres scales with the wave-function spectral weight projected to one unit cell at the top surface with Bi termination. The calculated Fermi level has been shifted about 130~meV upwards and the renormalization factor is set to be 1.77. Data were taken at 10K with circularly polarized photons of h$\nu$~=~64~eV.}
\label{theoretical calculations}
\end{figure}

The sketch of the crystal structure of PtBi$_2$ (P$\overline3$) is displayed in Fig. 1 (a) with lattice parameters: a=b=6.57 $\AA$ and c=6.16 $\AA$~\cite{PtBi2 crystal}. The structure is centrosymmetric with an inversion centre at the midpoint of the shortest Bi-Bi bond in the primitive cell. Along the $c$ axis, this bismuthide consists of alternative layers of Pt-Bi and Bi-Bi bonds, forming two possible cleavage planes. However, the bond length of Pt-Bi bonds is shorter than that of the Bi-Bi bonds, indicating higher bond energy in the Pt-Bi bonds. Therefore samples are most easily to be cleaved between the two Bi layers [marked by the blue dashed line in Fig. 1(a)]. Later we will show that the surface states probed by our photoemission experiments are more in line with the first-principles band structure slab calculation for the Bi-Bi plane, which further confirms the above judgement. The reduced BZ corresponding to this crystal structure is shown in Fig. 1(c).

%Fig4 kz
\begin{figure}[t]

\includegraphics[width=8.5cm]{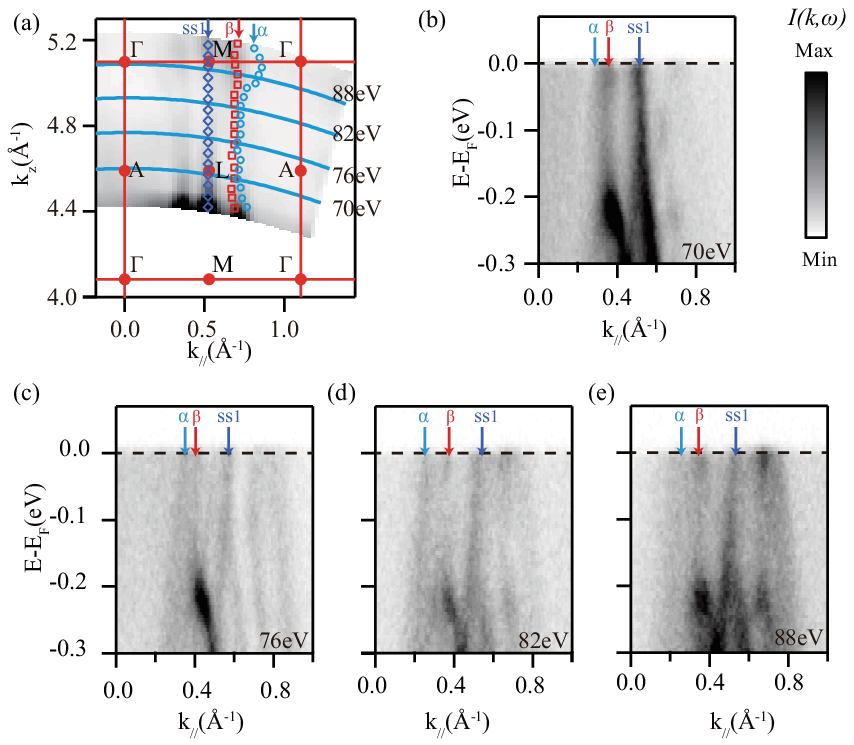}
\caption{(Color online) (a) The photoemission intensity of PtBi$_2$ in the $\Gamma$\emph{M}\emph{L}\emph{A} plane. The photoemission intensity map integrated over [$E_F$ - 10meV, $E_F$ + 10meV]. Different k$_z$'s were obtained by varying the photon energy as indicated by the blue solid lines, where an inner potential of 15eV is used to calculate k$_z$. The energy of right circularly-polarized photons ranges from 64 to 94~eV. (b)-(e) The photoemission intensities along or parallel to the $\Gamma-$\emph{M} direction, taken at different photon energies.}
\label{kz dependence}
\end{figure}

In Fig. 1(b), the X-ray diffraction (XRD) pattern of a typical PtBi$_2$ crystal only displays a series of (00n) peaks without any trace of secondary phases. Preliminary photoemission measurements reveal a clear valence band dispersion of PtBi$_2$ [Fig. 1(d)], confirming the high quality and good cleavage surface of our sample. Comparing with density functional theory (DFT) calculations, we can identify features between $-$5~eV and $-$3~eV as mainly contributed by the Pt 5\emph{d} states, while the spectral weight near E$_{F}$ is dominated by the Bi 6\emph{p} states. However, we note that the intensity analysis of EPMA gives the actual composition of Pt~:~Bi=1~:~1.838, well off the nominal stoichiometry, implying a substantial bismuth deficiency in this sample.

Next we choose 64 eV photons to probe the low-lying band structure of PtBi$_2$ in the $A$-$L$-$H$ $k_z$ plane assuming an inner potential of 15 eV as discussed below. Fig. 2(a) shows the photoemission intensity map taken with the circularly polarized photons by integrating photoemission intensities over a [$-$5meV, $+$5meV] window around $E_F$. In accord with the crystal structure, the resulting Fermi surface map shows a hexagonal configuration. Combined with the high-symmetry-direction photoemission intensity plots [Figs. 2(b) and 2(c), along the $A$-$L$ and $A$-$H$ directions, respectively], two bands assigned as $\alpha$ and $\beta$ which cross $E_F$ along the $A$-$L$ direction can be resolved. However, along the $A$-$H$ direction, while the band $\alpha$ crosses $E_F$ forming a large electron-like Fermi pocket around the $A$ point, the $\beta$ band only shows a hole-like feature with the band top at around -60 meV. In addition, around the $L$ point of the BZ there exists a Dirac-like band dispersion (assigned as ss1), the upper part of which forms a small electron pocket on the boundary of the BZ.

\begin{figure}[t]
\includegraphics[width=8.5cm]{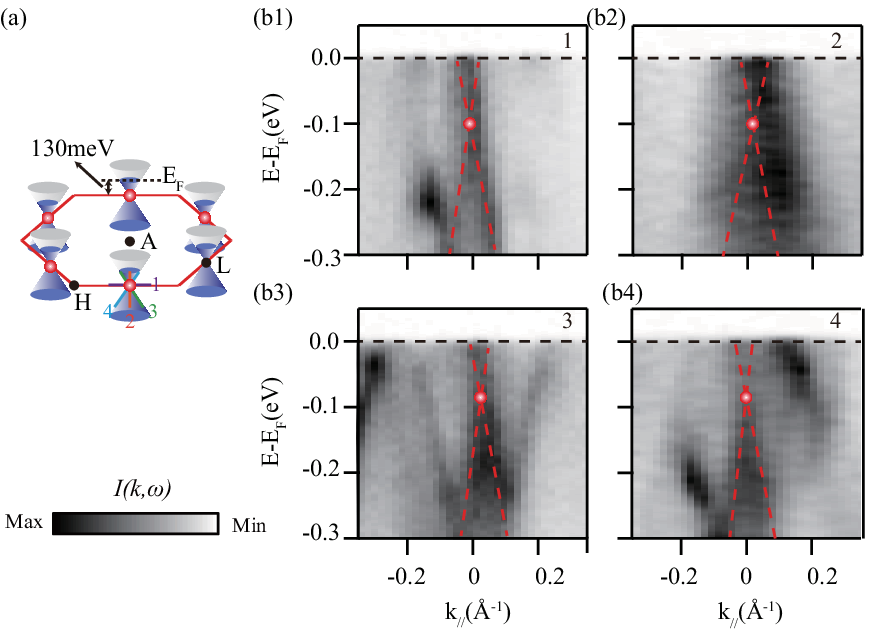}
\caption{(Color online) (a) The schematic diagram of a general Dirac cone. Dirac points, marked by red spheres, are located around -130~meV. Dispersions crossing the Dirac point show linear features. The grey plane represents the Fermi energy while the hexagon enclosed by black lines presents the corresponding Brillouin zone of PtBi$_2$. The different cut directions are indicated by the blue solid lines. (b1)-(b4) Photoemission intensity plots along cuts 1-4 respectively. Red dashed lines are guides for the eye, highlighting the linear dispersion around the Dirac point. Data were taken at 10K with circularly polarized photons of h$\nu$~=~64~eV.}
\label{Dirac point}
\end{figure}

The photon energy in our experiments ranges from 64~eV to 120~eV. Based on the universal mean-free-path curve for electron inelastic scattering in solids~\cite{universal curve}, the electron escape path is estimated to be about 5 $\AA$, corresponding to the first one or two layers in PtBi$_2$. Consequently, surface states are likely to be prominent in our photoemission experiment. To disentangle the complexity of the mixed bulk and surface states in the vicinity of $E_F$, we conducted detailed DFT band calculations for bulk and both the Pt-Bi and Bi-Bi cleavage planes. After careful comparison, we found that the Bi-Bi cleavage plane together with the bulk DFT calculations are more in line with our experimental data. In Fig. 3 we show the comprehensive characterization of the surface and bulk electronic structure of PtBi$_2$. The resulting band dispersion and second derivative plots together with colored guide lines along the $A$-$L$ and $A$-$H$ directions are presented in Figs. 3(a) and Figs. 3(b), respectively. The corresponding band structure calculations along two directions are displayed in Figs. 3(a3) and 3(b3) using the same color code. Here, a renormalization factor of 1.77 has been applied. From the good qualitative agreement with the data, we can further identify some more bulk and surface states in the vicinity of $E_F$, labeled by $\sigma$, $\eta$ and ss1, ss2, ss3 along the $A$-$L$ direction [Fig. 3(a1)]. Similarly, the bulk bands $\gamma$ and $\sigma$ and surface state ss5 can be resolved along the $A$-$H$ direction in Fig. 3(b1). %Although there is a qualitative agreement between experiment and the band structure calculations, at the quantitative level there are still some discrepancies.

Here, we note that calculated bands based on stoichiometric PtBi$_2$ should be shifted downward by around 130~meV to get a better match with the experiment. For example, the Dirac point of the calculated surface states around the $L$ point is close to $E_F$. However, our experiment shows that the energy of this Dirac point is actually located around $-$ 130~meV. This discrepancy might suggest electron doping induced by Bi vacancies.

%kz dependence
In Fig. 4, we show the result of photon energy dependence measurements on the Bi-Bi termination along the $\Gamma$ (\emph{A})-\emph{M} (\emph{L}) direction to comprehensively study the $k_z$ dependence of the electronic structure and thus further verify the above identification of surface states and bulk bands. Fig. 4(a) shows the photoemission intensity map in the $\Gamma$-$M$-$L$-$A$ plane. Here, we estimate all the $k_z$ values [illustrated by solid blue lines] according to the free-electron final-state model~\cite{Hufner}, where an inner potential of 15~eV was applied to obtain the best fit. The surface state ss1 can be clearly resolved as a straight, non-dispersive line extending over an entire BZ in the $k_z$ direction. Variation in the photoemission intensity might result from matrix element effects. The absence of any detectable $k_z$-dispersion indicates that this state is of two-dimensional nature. This finding can be further confirmed by the photoemission intensity plots along the $\Gamma$ (\emph{A})-\emph{M} (\emph{L}) direction taken with typical photon energies [Fig. 4(b-d)], in which the Fermi crossings of ss1 do not show any noticeable $k_z$ dependence. \textbf{In sharp contrast, both the Fermi-surface cross sections of the identified bulk $\alpha$ band and $\beta$ band and the corresponding Fermi crossings show evident out-of-plane energy dispersion, consistent with their three-dimensional nature. These findings further confirm our identification of surface states and bulk bands.}

%Dirac dispersion
Among all the identified bulk and surface states in the vicinity of $E_F$, the marked surface state ss1 with a massless Dirac dispersion at the boundary of the BZ naturally draws our attention. To investigate details of this feature, we have performed photoemission measurements along different cuts as illustrated in Fig. 5(a). Massless linear dispersions are observed along all these directions [Figs. 5(b1-b4)]. The experimental Fermi velocity of this linear band is $3.39\pm0.4\times10^5$ m/s, in good agreement with the calculated prediction of $3.77\times10^5$ m/s. These findings confirm the existence of the predicted surface Dirac cone around the BZ boundary near $E_F$.

Naturally, it is alluring to associate the anomalous large linear MR in PtBi$_2$ with such a linear surface band dispersion, in analogy to the case of topological insulators. However, we have carefully derived the $Z_2$ invariant $\nu_0$ for PtBi$_2$ to evaluate whether the observed surface state is topologically non-trivial, similar to three-dimensional topological insulators~\cite{FuliangPRB}. Our calculation proves that this surface state is topologically trivial, and this massless Dirac dispersion is accidentally caused by the band dispersion. Actually, for the Bi-Bi terminal surface, the $M$ point is located at a time-reversal-invariant point in the 2D BZ, and the states around $M$ are only inversion symmetric. Thus, once the surface band dispersion is expanded around this high-symmetry point, there must exist the first-order term, namely the massless linear dispersion. We note that such Dirac dispersion around the $M$ point is just allowed by the lattice symmetry but not necessarily topologically protected.

Consequently, this topologically trivial surface state is not protected by the time-reversal symmetry, and it may not survive the preparation process for transport measurements. Besides, different from the topological insulators, both the bulk carrier density and mobility of PtBi$_2$ are high, and the surface conductivity is expected to be small compared to the bulk. Thus, we conclude that the large linear MR can not solely caused by the quantum limit of the surface Dirac cones, which is consistent with the transport result that the possible quantum linear MR arising from the degenerate Dirac fermions in the quantum limit is not observed up to 15~T~\cite{Zhuan,Z. X. Shi}.

On the other hand, the classical disorder model can provide another explanation for the presence of large linear MR in solids~\cite{MR classical model,MR classical model2}. For PtBi$_2$, transport has shown that dMR/dB exhibits a temperature dependence identical to that of the Hall mobility and the $\frac{\mathrm{d} MR}{\mathrm{d} B}$ versus $\mu$ curve is fitted linear, strongly suggesting the classical disorder origin of the large linear MR in PtBi$_2$. Actually, both our EPMA measurement and the discrepancy between the nominal and the actual electron doping strongly suggest that there must exist a significant Bi deficiency. The resulting possible inhomogeneous carrier or mobility distribution would probably induce spatial fluctuations of the local current density in both magnitude and direction, and consequently lead to the large linear MR in PtBi$_2$~\cite{Zhuan}.

%---------------------------summary---------------------------------

\section{Conclusion}

To summarize, we have comprehensively characterized the surface and bulk electronic structure of PtBi$_2$ using ARPES measurements together with band structure calculations. We have identified the Bi-Bi cleavage plane and then distinguished the bulk from the surface bands on the resulting surface. Through careful comparison with calculations, we demonstrate the significant electron doping in this sample, which suggests a substantial Bi deficiency. Moreover, we have discovered a Dirac-cone-like surface state on the boundary of the BZ, which is identified as an accidental Dirac band without topological protection. Our findings exclude the linear band dispersion induced by the quantum limit but land support to the classical disorder model accounting for the unconventional large linear MR of PtBi$_2$.

\begin{center}
\textbf{ACKNOWLEDGMENTS}
\end{center}
We gratefully acknowledge the helpful discussions with Dr. Wei Li. This work is supported by the National Key R\&D Program of the MOST of China (Grant No. 2016YFA0300200) and the National Science Foundation of China (Grant Nos. 11274332, 11574337, 11227902 and U1332209). D. W. S. is also supported by the ¡°Strategic Priority Research Program (B)¡± of the Chinese Academy of Sciences (Grant No. XDB04040300) and the ¡°Youth Innovation Promotion Association CAS¡±.

\end{document}